\documentclass[allclo]{FBSart}
\usepackage{amsfonts}
\usepackage{amssymb}
\usepackage{bm} 
\usepackage{graphicx}
\usepackage{amsmath}
\usepackage{slashed}
\def\bea{\begin{eqnarray}}
\def\eea{\end{eqnarray}}
\def\bal{\begin{align}}
\def\eal{\end{align}}

\title{High-precision covariant one-boson-exchange potentials for $np$ scattering below 350 MeV}
\author{\vspace{-5mm}Franz Gross\instnr{1,2}, Alfred Stadler\instnr{3,4} \vspace{-5mm}}
\instlist{Thomas Jefferson National Accelerator Facility, Newport News, VA 23606 
\and College of William and Mary, Williamsburg, VA 23187
\and Centro de F\'isica Nuclear da Universidade de Lisboa, 1649-003 Lisboa, Portugal 
\and Departamento de F\'isica da Universidade de \'Evora, 7000-671 \'Evora, Portugal}

\runningauthor{Franz Gross and Alfred Stadler}
\runningtitle{High-precision covariant one-boson-exchange potentials for $np$ scattering}
\begin{document}

\maketitle
\vspace{-5mm}
\begin{abstract}
Using the Covariant Spectator Theory (CST), we have found One-Boson-Exchange (OBE) potentials that fit the 2006 
world $np$ data below 350 MeV with a $\chi^2/N_\mathrm{data}$ very close to 1, for a total of 3788 data. 
Our potentials have significantly fewer adjustable parameters than previous high-precision potentials, and they also 
reproduce the experimental triton binding energy without introducing additional irreducible three-nucleon forces.
\end{abstract}

The OBE mechanism played an important role in the long history of attempts to understand
the nucleon-nucleon (NN) interaction. It is still the dominant feature in modern realistic potential models which
describe the experimental observables with $\chi^2/N_\mathrm{data} \approx 1$, such as the Nijmegen \cite{Sto94}
and CD-Bonn \cite{Mac01} potentials. However, in order to achieve such a good fit, these potentials abandoned a 
pure OBE form and made several boson parameters partial-wave dependent, thereby increasing the number of 
adjustable parameters significantly. For instance, the best pure OBE potential of the Nijmegen group reached 
$\chi^2/N_\mathrm{data} = 1.87$ with 15 parameters  (Nijm93), while the impressive result of 
$\chi^2/N_\mathrm{data} = 1.03$ was obtained at the expense of 41 parameters (Nijm I).  
The Argonne group incorporated OBE only in the case of the pion, 
but also motivated their construction of otherwise largely phenomenological realistic 
potentials like AV18 by the 
apparent failure of the OBE mechanism to allow a perfect fit to the data \cite{Wir95}. 

We found that within the CST it is, in fact, possible to derive realistic OBE potentials, 
and that these require comparatively few parameters. This somewhat surprising finding contradicts the earlier 
conclusion and common belief that the OBE mechanism is missing some important feature of the $NN$ interaction.

In CST, the $NN$ scattering amplitude is obtained from a covariant integral equation with a very similar structure 
to the  nonrelativistic Lippmann-Schwinger equation, but with
a covariant kernel in the place of the potential.  A major difference, however, is the appearance of negative-energy 
nucleon states.

Previous CST models of the kernel, such as the models of \cite{Gro92} and the updated versions including off-shell
couplings of scalar mesons used in \cite{Sta97}, 
had been obtained by fitting the potential parameters to the Nijmegen \cite{Stoks:1993tb} or 
VPI \cite{SAID} phase shifts. Only in a second step the $\chi^2$ to the observables was determined. 
In contrast, the models presented here were fitted {\em directly to the data}. 
We found this a significant improvement, because the best fit to the Nijmegen 
1993 phase shifts did not yield the best fit to the 2006 data base.

\begin{table}
\caption{Comparison of precision $np$ models and the 1993 Nijmegen phase shift analysis.  Our calculations are in bold face.}
\label{tab:1}
\begin {tabular}{lcc|ccc} \multicolumn{3}{c}{models}&\multicolumn{3}{c}{$\chi^2/N_\mathrm{data}$}\cr
\hline
Reference & $N_\mathrm{par}$ & max year+1 & $\;\;$1993$\;\;$ & 2000& 2007 \cr
\hline
PWA93\cite{Stoks:1993tb} &39 ($pp$ and $np$) &1993 &  0.99 & --- & ---\cr
&&&{\bf 1.09} &{\bf 1.11}&{\bf 1.12}\cr
Nijm I\cite{Sto94}&41& 1993 &1.03 &---&---\cr
AV18\cite{Wir95} & 40 & 1995 &1.06 & --- & --- \cr
CD-Bonn\cite{Mac01} & 43 & 2000 & --- &1.02&--- \cr
WJC-1& 27$\;$ & 2007 &{\bf 1.03}& {\bf 1.05} & {\bf 1.06} \cr
WJC-2 & 15$\;$ & 2007 & {\bf 1.09}& {\bf 1.11} & {\bf 1.12}
\end{tabular}
\vspace{-0.2in}
\end{table}

Our best new model, WJC-1, is based on the exchange of one isoscalar and one isovector meson in each of the 
pairs of
pseudoscalar ($\eta$, $\pi$), scalar ($\sigma_0$, $\sigma_1$), vector ($\omega$, $\rho$), and axial vector 
($h_1$, $a_1$) bosons. These exchanges already provide the most general 
spin-isospin structure of an $NN$ kernel, at least when the vector mesons have Dirac and Pauli couplings and 
external nucleons are on-shell. Except for the pions, the other boson masses are determined through the fits. 
The axial vector mesons are comparatively heavy, and we decided to treat them as contact interactions. 
Charge symmetry is broken by treating charged and neutral pions 
independently, and by adding a simplified one-photon-exchange interaction. The pions couple to nucleons through
a mixture of PV and PS coupling.
WJC-1 has the comparatively low number of 27 adjustable parameters. In order to better compare to older OBE 
potentials, like Nijm93, we constructed model WJC-2, which eliminates axial vector mesons and a number of other 
degrees of freedom, with only 15 adjustable parameters remaining.
Details about the structure of the 
interaction models and the values of their parameters can be found in Ref.\ \cite{Gro07}. 

Table 1 shows the $\chi^2/N_\mathrm{data}$ for our models, in comparison with previous 
high-precision potentials and with the Nijmegen phase shift analysis. It also shows, through the 
cut-off year for data included in the fits, how
the $\chi^2$ increases with the growing database. Note that our result for PWA93 with the 
1993 database is higher than the one of Ref.\ \cite{Stoks:1993tb} because our database includes 3010 data 
prior to 1993 versus 2514 data used in the PWA93 fit.

The phase shifts of WJC-1 and WJC-2 are displayed in Figure 1, together with the Nijmegen phase shifts.
While their overall behavior is very similar, in some cases differences of more than one degree occur, which 
may be significant.

\begin{figure}
\vspace{-5mm}
\centerline{\includegraphics[width=14cm]{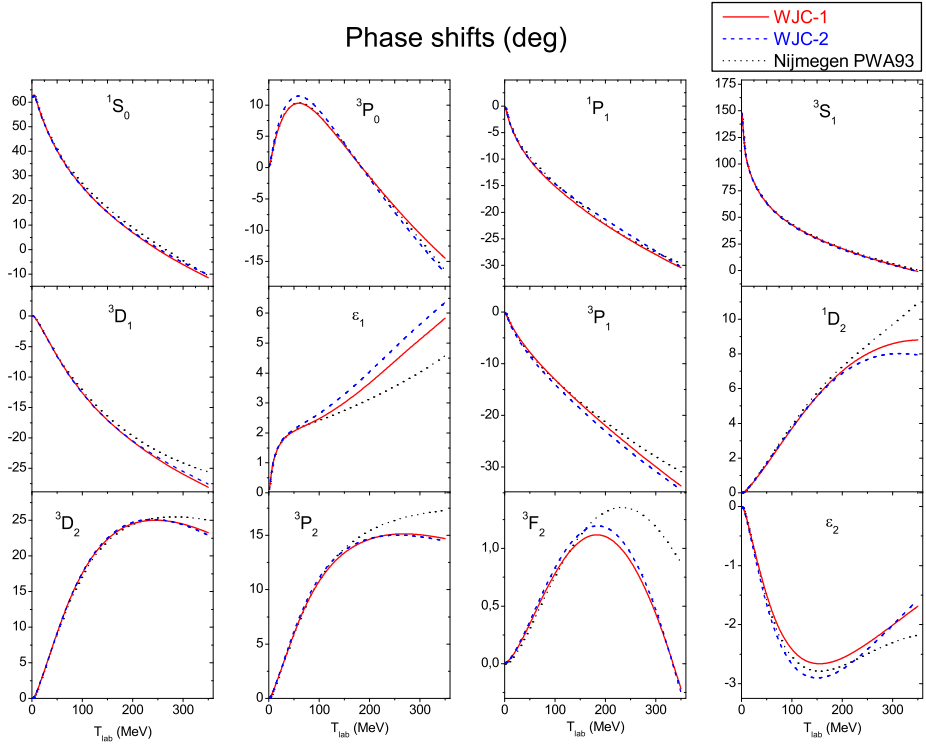} }
\vspace{-10mm}
\caption{Phase shifts of $np$ scattering with models WJC-1 (solid lines), WJC-2 (dashed lines), and the Nijmegen 
PWA of 1993 (dotted lines).}
\vspace{-5mm}
\end{figure}

We had already found in calculations with previous CST models that the $\sigma NN$ off-shell coupling 
strongly influences the triton binding energy \cite{Sta97}, and that the best fit to the $NN$ data automatically 
also leads to the correct triton binding energy without additional three-body forces. It is remarkable to find this to be the case again, for {\em both} 
WJC-1 ($E_t=-8.48$ MeV) and WJC-2 ($E_t=-8.50$ MeV), which are after all quite different.

We conclude that the OBE concept, at least in the context of the CST where it can be comparatively easily 
extended to the treatment of electromagnetic interactions and systems with $A>2$, can be a very effective 
description of the nuclear force.
Model WJC-1 provides also a new phase shift analysis, updated for 
all data until 2006, which is useful even if one does not work within the CST.

\begin{acknowledge} F.\ G.\ was supported by Jefferson Science Associates, LLC under 
U.S. DOE Contract No.~DE-AC05-06OR23177. A.\ S.\ was supported by FCT under grant No.~POCTI/ISFL/2/275. 
\end{acknowledge}

\end{document}